\documentclass[a4paper, amsfonts, amssymb, amsmath, reprint, showkeys, nofootinbib,superscriptaddress,onecolumn ]{revtex4-2}
\usepackage{float}
\usepackage{graphicx}
\usepackage{longtable}
\usepackage[left=15mm,right=13mm,top=23mm,columnsep=15pt]{geometry} 
\usepackage[caption=false]{subfig}
\usepackage{hyperref} 
\usepackage{multirow}
\usepackage{array}
\usepackage{appendix}
\usepackage{color}

\newcolumntype{L}{>{\centering\arraybackslash}m{6cm}}

\begin{document}
\title{Hyperon productions from Au+Au collisions at $\sqrt{s_{NN}}$=200 GeV.}

\author{Purabi Ghosh}
    \affiliation{Department of Applied Physics and Ballistics, Fakir Mohan University, Balasore-756019, India.}
\author{Sushant K. Singh}
    \affiliation{Variable Energy Cyclotron Centre, 1/AF, Bidhan Nagar , Kolkata-700064, India}
    \affiliation{HBNI, Training School Complex, Anushakti Nagar, Mumbai 400085, India}
\author{Santosh K. Agarwalla}
    \affiliation{Department of Applied Physics and Ballistics, Fakir Mohan University, Balasore-756019, India.}
\author{Jajati K. Nayak}
    \email[Correspondence email address: ]{jajati-quark@vecc.gov.in}
    \affiliation{Variable Energy Cyclotron Centre, 1/AF, Bidhan Nagar, Kolkata-700064, India}

\date{\today} 

\begin{abstract}
We evaluate centrality dependence of hyperon, $\Lambda, \Xi, \Omega$ and $K$ meson yields from Au+Au collisions at $\sqrt{s_{NN}}$=200 GeV using rate equation and compare with experimental observations. An increase in the yield per participating nucleon is observed with centrality. Higher rate of multistrange productions compared to single-strange are also observed in case of  central collisions. Using rate equation or momentum integrated Boltzmann equation we discuss the yield microscopically which are then normalised with thermal pions to get the {\it yield ratio},($H_s+\bar{H_s}$)/($\pi^++\pi^-$) at various centralities. We have also evaluated the relative {\it yield of anti hyperons per participant} at various centralities and compared with observations made at RHIC. Increase in the rate of $\Xi$ yield compared to $\Lambda$ is realised in most central collisions. We find multi strange $\Xi,\Omega$ freeze-out at a temperature very close to $T_c$ and $\Lambda, K$ do little later.
\end{abstract}

\maketitle 
\section{Introduction}
Strange hadrons are sensitive probes to understand the hot and dense medium formed in relativistic heavy ion collisions. Enhancement of strangeness productions in heavy ion central collisions relative to $p-p$ collisions was argued long back as a signal of quark gluon plasma formation\cite{MG04,rafelski82}. With similar argument the enhancement is supposed to increase with the increase in strangeness content of the hadron. This is why, the study on multi-strange baryons and their anti-particle productions in heavy-ion and $p-p$ collisions at Relativistic Heavy Ion Collider(RHIC) and Large Hadron Collider(LHC) are very much contemporary. Recently, the yield of $\Lambda, \ \Xi,$ and $\Omega$ have been measured from both central and peripheral Au+Au collisions at $\sqrt{s_{NN}}$=200 GeV at midrapidity \cite{adam_star_prl07}. An increase in the yield of hyperons with number of participating nucleons, $N_{part}$ is observed from peripheral to central collisions. The yield per participant (yield  normalised with $N_{part}$) is also observed to increase gradually. The ``{\it yield per participant}'' is an important observable as it reflects the formation probability of a hadron from the bulk of the matter. This increase in {\it yield per participant} is more for multi-strange hyperons $\Xi$ and $\Omega$ than the single strange hyperon $\Lambda$ as reported in \cite{adam_star_prl07}. 


As the size of the colliding system increases, number of participants increases. This leads to more number of interactions and we observe an increase in the hadronic yield. But when we consider {\it yield per participant} as observable, then we try to remove the system size dependance. Then, it is expected that the {\it yield per participant} should go flat if the relationship between yield and $N_{part}$ is linear. However, from the measurements of anti-hyperons, it is not observed\cite{adam_star_prl07}. {\it Yield per participant} is enhanced when we move from peripheral to central collisions. This indicates a non-linear dependance of yield with $N_{part}$. The rate of enhancement is more for multistrange baryons compared to single-strange baryons $\Lambda$. 

Analogous to the above argument, another observable {\it yield ratio} ($H_s+\bar{H_s}$)/($\pi^++\pi^-$) provides similar observations from $Pb-Pb$ collisions at $\sqrt{s_{NN}}$= 2.76 TeV. $H_s$, here, represents the yield of strange hadrons. In one of the recent measurements, ALICE Collaboration has reported the yield of strange mesons($K^0_s$) and hyperons ($\Lambda^0, \Xi^-$ and $\Omega^-$) at various centralities from $Pb-Pb$ collisions at $\sqrt{s_{NN}}$=2.76 TeV \cite{multistrange_alice_plb14,alice_abelev_prl12}. The yield of strange hadrons (with its anti-particle) obtained from $p-p$, $p-Pb$ and $Pb-Pb$ collisions are normalised with charged pions ($\pi^++\pi^-$) and displayed as a function of multiplicity in recent article \cite{alicenature17}. Authors in \cite{alicenature17} take the measurements of $Pb-Pb $ collisions (at $\sqrt{s_{NN}}$=2.76 TeV) from \cite{multistrange_alice_plb14,alice_abelev_prl12} and  $p-p$ collisions (at $\sqrt{s_{NN}}$=7 TeV) from \cite{abelev_ALICE_plb12,acharya_alice_prc19,gyula_alice} and $p-Pb$ collisions (at $\sqrt{s_{NN}}$=5.02 TeV) from \cite{alice_adam2016}. The normalised yield of strange hadrons ($H_s+\bar{H_s}$)/($\pi^++\pi^-$), conveniently called as '{\it yield-ratio}' show a smooth rising pattern with centrality(or multiplicity) for the hyperons and strange mesons following a saturation towards higher multiplicities. Observations from both {\it yield per participant} and {\it yield ratio} pose an important question regarding hyperon productions-whether the multistrange production yield is enhanced due to the increase in number density of constituent quarks or not? 

Along this line, one more observation has been reported \cite{multistrange_star_prc08} where differences in production rates of strange and multistrange baryons have been observed in Au-Au
compared to $p-p$ collisions at same energy, $\sqrt{s_{NN}}$= 200 GeV. When the hyperon yields are scaled down by the number of participants(participating nucleons) an enhanced yield is observed relative to p+p collisions. The observed enhancement increases with the strangeness content of the hyperon, and that happens with centrality for all the hyperons. 

In this article, we discuss hyperon productions at $\sqrt{s_{NN}}$= 200 GeV, RHIC energy with a microscopic description using rate equation. We evaluate both {\it yield ratio} and {\it relative yield per participant} of hyperons for various centralities and compare with the observations from Au+Au collisons at midrapidity ($|\eta| <0.5$)\cite{adam_star_prl07}. We also discuss the freeze out properties of strange hadrons by analyzing these two observables. From the analysis within the ambit of our model, we try to comment on the probable reason for more multi strange productions compared to single strange in central collisions .

As far as single strange hadron ($K$ and $\Lambda$) productions are concerned, several literatures are available\cite{rafelski82,kapusta86,cugnonprc90,LR08,RL99,BT07,sgupta10,CORS,MG04,andronic06,jknacta06,jknprc10,tawfik09,nayak2011npa} in recent past, but these articles do not discuss multistrange productions. There are some literatures where the authors discuss subthreshold multistrange $\Xi$ productions \cite{steinheimerjpg16} with the inclusion of new resonance decay channels from massive baryons. A transport calculation on multistrange productions at SPS energy using ultra relativistic quantum molecular dynamics(UrQMD) is available ~\cite{bassplb99}. Authors in \cite{kolomeitsev12,kolomeitsev15} use minimal statistical hadronisation model to explain NICA data. Present calculation adopts a microscopic approach and it discusses multi strange productions along with single strange hadrons at $\sqrt{s_{NN}}$=200 GeV, RHIC energy.

The manuscript is presented in the following manner. The rate equation for yield calculation is described in the Sec.\ref{sec:evolution}. Equations for hyperons and strange mesons along with evolution of the bulk is discussed in this section. The interactions of various hadronic species and their cross sections which tell about the microscopic origin of hyperon yields are considered in Sec.\ref{sec:interaction}. The rates of production are also described here. Due to expansion of the system, the temperature ($T$) and chemical potential ($\mu$) of the bulk change with time. The evolution of $T$ \&,$\mu$ are also described in this section. Theoretical estimation and its comparision with data are discussed in Sec.\ref{sec:results}. Finally, we summarize in Sec.\ref{sec:summary} with conclusion.
\section{\label{sec:evolution} Yield of hyperons using rate equation}
There are several evidences which suggest that an initial hot-dense partonic system is produced in Au-Au collisions at RHIC at $\sqrt{s_{NN}}$=200 GeV\cite{rhicwhitepaper05brahms,rhicwhitepaper05phobos,rhicwhitepaper05star,rhicwhitepaper05phenix}. It is also known by conventional wisdom that the hot-dense partonic system converts into hadronic system during the evolution as temperature falls below $T_c$. In the hadronic system, the hadrons change their numbers continuously due to rescatterings, till the chemical decoupling happens at a temperature $T_{ch}$, known as the chemical freeze-out temperature. At $T_{ch}$, the hadron yields get fixed and do not change subsequently. With subsequent colling, the hadrons decouple kinetically at a temperature $T_{KF}$, called as the knietic freeze-out temperature. After the knietic freeze-out, while particles stream freely towards the detector, some of the hadronic resonance states decay which is not a part of the rescattering effect of the medium. The chemical freeze-out temperatures may be same or different for different hadronic species. If $T_{ch}$ is same for all species, then this is called as common freeze-out(chemical) temperature, and if $T_{ch}$ is different for different species, then such a scenario can be descibed through sequential freeze-out mechanism. Sequential freeze-out is recently discussed in the following article ~\cite{bellewiedepj18}. Whether the freeze-out is common or sequential, it depends on the species of particle and the dynamical situation of the hadronic system. 

To understand the probable freeze out scenario of hyperons and Kaons at RHIC energy, we here analyse the experimental observations by evaluating $K, \Lambda, \Xi,\Omega$ yield using rate equation or momentum intergated Boltzmann equation. We assume non-strange hadrons to provide a thermal background to the strange hadrons in the system. The use of rate equation is relevant here as the hadronic system produced in relativistic nuclear collisions at RHIC/LHC is supposed to be a dilute gas. Similar formalism has been used by us to explain the strange hadron production at LHC energy in \cite{ghoshprd20}. Bjorken expansion is considered for the evolution of the bulk as it is relatively easier to analyze analytically. Again, our focus here, is to look at the ratio of the numbers (ratio of integrated yields) at freeze-out. The qualitive behaviour would be unchanged if we use 3-dimensional expansion to the present analysis. The value of freeze-out parametres may change. The results of the study of observables like flow and $p_T$ spectra of identified hadrons would have been different with 3- and 1-dimensional expansion . A relevant 3-dimensional Hubble like expansion in hadronic phase is under progress to compare the results with Bjorken expansion and would be communicated later.


The complete set of rate equations for a system with hyperons, $\Lambda, \Xi, \Omega$, and strange mesons, $K, \bar{K}$ has been documented in Ref.~\cite{ghoshprd20}. We solve the same set of equations to analyze the strange hadron yields at RHIC for $\sqrt{s_{NN}}$=200 GeV. The set consists of first-order coupled ordinary differential equations which describe the time evolution of hadronic yields due to interactions within the system in presence of a dilution term due to expansion. The important inputs which are needed to solve the equations are the hadronic cross-sections, which are mentioned in the next section.  To discuss the rate of production for a particular reaction, one may think of a channel $a+b\rightarrow c+d$ where the thermal average rate $\langle \sigma v\rangle_{ab\rightarrow cd}$ can be described as below. Assuming Maxwell-Boltzmann statistics for the particles, the reaction rate at a temperature $T$ is given by~\cite{kapusta86,gondolo91}
\begin{align}
 \langle \sigma v\rangle_{ab\rightarrow cd} &=\frac{T^4}{4}\mathcal{C}_{ab}(T)\int _{z_0}^{\infty} \, dz\, [z^2-(m_a/T+m_b/T)^2]\nonumber \\
 &\times [z^2-(m_a/T-m_b/T)^2]\sigma K_1(z),
  \label{eqn_reacrate}
\end{align}
where 
$$\mathcal{C}_{ab}(T) = \frac{1}{m_a^2m_b^2K_2(m_a/T)K_2(m_b/T)}.$$ 
Here $\sigma$ and $v$ respectively denote the scattering cross section and relative M\"{o}ller velocity of incoming particles $a$ and $b$ for the process $ a+b \rightarrow c+d$. $K_2$ is the modified bessel function of second kind. $z$ is basically $E/T$, where $E$ is the total centre of mass energy. The lower limit of the integration is $z_0=\text{max}(m_a+m_b,m_c+m_d)/T$. For a detailed derivation of the rate equations, one can see Appendix A in~\cite{ghoshprd20}. 
\begin{table*}
 \centering 
 \caption{Integrated yields of $\pi$ \cite{abelev_star_prc09} and $K_s^0$ \cite{agakishiev_star_prl12} at midrapidity $|y|<0.5$ for various centralities from Au+Au collisions at $\sqrt{s_{NN}}$=200 GeV which are used to calculate the {\it yield ratio} for $K^0_s$ and the same is tabulated here. }
 \begin{tabular}{ |c|c|c|c|c|c|c|} 
  \hline
  Centrality & $N_{part}$ & $dN/dy$ & $dN/dy$& $dN/dy$ & & \\
  (\%) & \cite{agakishiev_star_prl12} &  $\pi^+$\cite{abelev_star_prc09}& $\pi^-$\cite{abelev_star_prc09}& $K^0_s$\cite{agakishiev_star_prl12} & $2K^0_s/(\pi^++\pi^-)$ & $K^0_s/{N_{part}}$ \\
    \hline
   0-5 & 350 $\pm$4  & 322$\pm$25 & 327$\pm$25  & 43.5$\pm$2.4 &0.134$\pm$0.0052 & 0.124$\pm$0.007 \\
  \hline
   10-20 & 238$\pm$5 & 194$\pm$15  & 196$\pm$15  & 27.80$\pm$1.4 & 0.142$\pm$0.005  & 0.116$\pm$0.006  \\ 
   \hline
   20-40 & 147$\pm$4 &112.1$\pm$12.09 & 112.8$\pm$12.09  & 16.50$\pm$0.83 & 0.146$\pm$0.0066 & 0.112$\pm$0.006  \\
  \hline
   40-60 & 67.5$\pm$2.7  & 47.5$\pm$5.25  & 47.6$\pm$5.30 & 7.26$\pm$0.49 & 0.152$\pm$0.007  & 0.107$\pm$0.008  \\
  \hline
   60-80 & 23$\pm$1.2 & 15.9$\pm$1.79 & 16$\pm$1.79 & 2.14$\pm$0.19  & 0.134$\pm$0.007 & 0.093$\pm$0.009  \\
  \hline
 \end{tabular}
 \label{table_data_k0s} 
\end{table*}
\begin{table*}
 \centering 
 \caption{Integrated yields of $\Lambda$,$\Xi$ and $\Omega$ at midrapidity $|y|<0.5$ for various centralities from Au+Au collisions at $\sqrt{s_{NN}}$=200 GeV are taken from \cite{adam_star_prl07}. The {\it yield ratio} for hyperons have been calculated and tabulated here. }
 \begin{tabular}{ |c|c|c|c|c|c|c|c|c|c|c|c|c|} 
  \hline
  Centrality & $N_{part}$ & $dN/dy$ & $dN/dy$& $dN/dy$ &$dN/dy$ &$dN/dy$ & & & & & & \\
  (\%) & \cite{adam_star_prl07}&  $\Lambda$\cite{adam_star_prl07}& $\bar{\Lambda}$\cite{adam_star_prl07}& $\Xi^-$ \cite{adam_star_prl07}& ${\bar\Xi^+}$\cite{adam_star_prl07} & $\Omega^-+\bar{\Omega^+}$ \cite{adam_star_prl07}&$\frac{(\Lambda+\bar{\Lambda)}}{(\pi^++\pi^-)}$ &$\frac{(\Xi^-+\bar{\Xi^+)}}{(\pi^++\pi^-)}$&$\frac{(\Omega^-+\bar{\Omega^+})}{(\pi^++\pi^-)}$&$\frac{\Lambda+\bar{\Lambda}}{N_{part}}$&$\frac{\Xi^-+\bar{\Xi^+}}{N_{part}}$&$\frac{\Omega^-+\bar{\Omega^+}}{N_{part}}$\\
    \hline
   0-5 & 352 & 16.7 & 12.7 & 2.17 & 1.83 & 0.53  & 0.0453 & 0.0061 & 0.00081  & 0.0835 & 0.01136 & 0.0015 \\
    & $\pm$3 & $\pm$1.12  & $\pm$0.92 & $\pm$0.19 & $\pm$0.21 & $\pm$0.056  & $\pm$0.003 & $\pm$0.0005 & $\pm$0.00009  & $\pm$0.004 & $\pm$0.0008  & $\pm$0.00015 \\
  \hline
   10-20 & 235 & 10.0 & 7.7 & 1.41  & 1.14 & - & 0.0454 & 0.0065 &- & 0.0753 & 0.0108 & -  \\
    & $\pm$9 & $\pm$0.71 & $\pm$0.51 & $\pm$0.09  & $\pm$0.09 & - & $\pm$0.003 & $\pm$0.00048 & - & $\pm$0.004 & $\pm$0.0068 &-  \\
   \hline
   20-40 & 141 & 5.53  & 4.30  & 0.72 & 0.62 & 0.17 & 0.0437 & 0.0059 & 0.00075 & 0.0697 & 0.0095 & 0.0012 \\
    & $\pm$ 8 & $\pm$0.39  & $\pm$0.05  & $\pm$0.028  & $\pm$0.036 & $\pm$0.022 & $\pm$0.0037 & $\pm$0.0005 & $\pm$0.00011 & $\pm$0.0048 & $\pm$0.0006 & $\pm$ 0.00017\\
  \hline
   40-60 & 62 & 2.07  & 1.64  & 0.26 & 0.23 & 0.063 & 0.039 & 0.0051 & 0.00066 & 0.0598 & 0.0079 & 0.0010 \\
    & $\pm$9 & $\pm$ 0.143  & $\pm$0.114  & $\pm$0.022 & $\pm$0.022 & $\pm$0.0089 & $\pm$0.0036 & $\pm$ 0.0005 & $\pm$0.0001 & $\pm$0.009 & $\pm$0.0012 & $\pm$0.0002  \\
  \hline
   60-80 & 21 & 0.58 & 0.48 & 0.063 & 0.061 & - & 0.0331 & 0.0038 & - & 0.05 & 0.0059 & -  \\
    & $\pm$6 & $\pm$0.041 & $\pm$ 0.032 & $\pm$ 0.005 & $\pm$0.0044 & - & $\pm$0.003 & $\pm$0.0003 & - & $\pm$0.014 & $\pm$ 0.001 & -  \\
  \hline
 \end{tabular}
 \label{table_data_hyperon} 
\end{table*}

Unlike Hadron Resonance Gas model, we obtain the yield by solving the rate equations. We assume the initial densities to be slightly away from equilibrium. In this study, we do not consider the initial QGP phase. The initial number densities ($n_i(T_i)$, with $T_i$ the initial temperature) of different hadrons can be constrained if the hadronization mechanism were known. Instead, we treat the initial number densities as parameters of the rate equations. Hence this calculation does not prohibit the initial QGP phase. Considering a Bjorken expansion of the hadronic medium, the temperature evolution is obtained numerically along with the rate equations. This is coded as Strange Hadron Transport in Heavy Ion Collision (SH-THIC). Evolution of baryonic chemical potential, $\mu_b$, is also considered, even though $\mu_b$ is small ($\sim$ 25 MeV) at RHIC, $\sqrt{s_{NN}}$=200 GeV. Baryonic chemical potential is obtained from following equation for baryon number conservation:
\begin{equation}
\partial_{\mu}n_b^{\mu}=0,
\end{equation}
where $n_b^{\mu}=n_b u^\mu$ denotes the net-baryon density flow vector, and which is a function of $T$ and $\mu_b$. If the expansion is considered to be along the $z$-direction, then the flow velocity, $u^\mu$, becomes $u^\mu=(\gamma,0,0, \gamma {v_z})$. We adopt the following definition for the net-baryon density 
$$n_b= \sum_{B=N,\Lambda,\Sigma,\Xi,\Omega}(n_B-n_{\bar B})$$
and ignore the contribution from massive baryons above $\Delta$ mass. For the equation of state, we take $p=c_s^2 \epsilon$. Now using the energy-momentum conservation equation, $\partial_{\mu}T^{\mu\nu}=0$~\cite{bjorken}, we can obtain the temperature evolution as
$${T^{\frac{4}{(1+c_s^2)}}\tau}=\text{constant}.$$
The initial baryon number densities $n_b^i$, time $\tau_i$ and temperature $T_i$ are parametres. The hadronic system evolution starts from $T_c$.

\begin{figure*}
\centering
\includegraphics[width=0.44\textwidth,height=6.5cm]{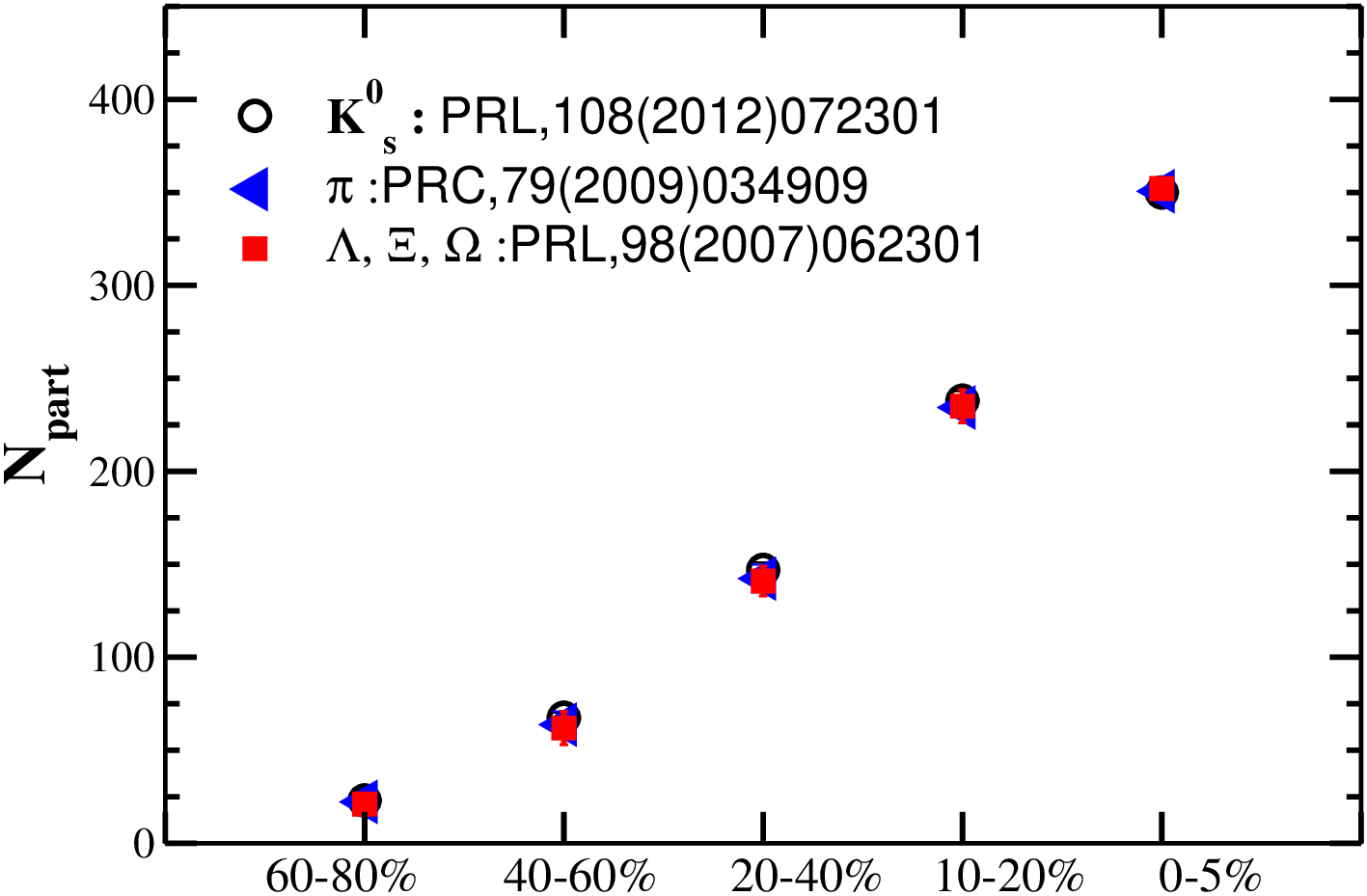}
\caption{$N_{part}$ vs centrality as reported in three references~\cite{adam_star_prl07,agakishiev_star_prl12,abelev_star_prc09}. }
\label{fig_npart_centrality}
\end{figure*}
\section{\label{sec:interaction} Scatterings of hadrons producing hyperons and strange mesons and cross section of various processes.}
Various scattering processes (along with inverse reaction) considered in our calculation to produce strange hadrons are:\\
$\pi \pi \leftrightarrow K \bar{K}$, $\pi \rho \leftrightarrow K \bar{K}$, $\pi N \leftrightarrow \Lambda K$, $\pi \Lambda \leftrightarrow \bar{K} N$, $\pi \Sigma \leftrightarrow \bar{K} N$, $\pi \Xi \leftrightarrow \bar{K} \Lambda$, $\pi \Xi \leftrightarrow \bar{K} \Sigma$, $\rho \rho \leftrightarrow K \bar{K}$, $\rho N \leftrightarrow \Lambda K$, $\bar{K}N \leftrightarrow K \Xi$, $p\bar{p}\leftrightarrow K^- \bar{K^+}$, $p\bar{p} \leftrightarrow \Lambda \bar{\Lambda}$, $p\bar{p}\leftrightarrow \Sigma^- \bar{\Sigma^+}$, $p\bar{p}\leftrightarrow \Omega \bar{\Omega}$, $p\bar{p} \leftrightarrow \Xi \bar{\Xi}$, $\Lambda\Lambda \leftrightarrow N \Xi$, $\Lambda\Sigma \leftrightarrow N \Xi $, $\Sigma\Sigma \leftrightarrow N \Xi$, $\Lambda \bar{K}\leftrightarrow \Omega^{-} K^0$, $\Sigma^{0} \bar{K}\leftrightarrow \Omega^{-} K^0$.\\

The symbols take their as usual meaning. Isospin combinations are considered for all those processes where specific isospin channels are not mentioned. A few of the above processes, $YY\rightarrow N\Xi$, $\bar{K} N\rightarrow K\Xi$ and $\bar{K} Y\rightarrow \pi\Xi$, $\bar{K} N\rightarrow \pi Y$, although mentioned in literature, but are not considered due to unreasonable higher value of theoretical cross-sections, which are also not verified experimentally. In the above description, $Y$ denotes either $\Lambda$ or $\Sigma$ hyperons. The details of the cross sections are mentioned in the appendix in a tabular form and also available in \cite{ghosharxiv19,ghoshprd20}. The cross-sections for the inverse processes are calculated using the principle of detailed balance\cite{cugnon84} as follows:
\begin{equation}
 \sigma_{f\rightarrow i}=\frac{{P_i}^2}{{P_f}^2}\frac{g_i}{g_f}\sigma_{i\rightarrow f},
\end{equation}
where $P_i, P_f$ are the momenta of incoming and outgoing channels in the center of mass frame, and  $g_i, g_f$ are the degeneracies, respectively. With above formalism the yields of strange hadrons, $K,\Lambda, \Xi, \Omega$ are calculated and presented in the next section.  

\section{\label{sec:results} Results}
Taking cross sections as input, the number densities of $K, \Lambda, \Xi, \Omega$ are evaluated for various centralities through the rate equations. Then the densities are normalised with thermal pion densities. Initial number density, $n_i$, initial temperature, $T_i$ and velocity of sound $c_s$ are various parameters in this calculation. The hadron phase starts at $T_c$ and the value is taken to be 156 MeV referring to the first principle QCD calculations based on lattice frame work ~\cite{swagato17}. We take $c_s^2$=1/5 for all cases. Different $n_i$ and $T_{ch}$ are taken for various scenarios to be discussed later.

\begin{figure*}
\centering
\subfloat{
   \includegraphics[width=0.44\textwidth,height=6.5cm]{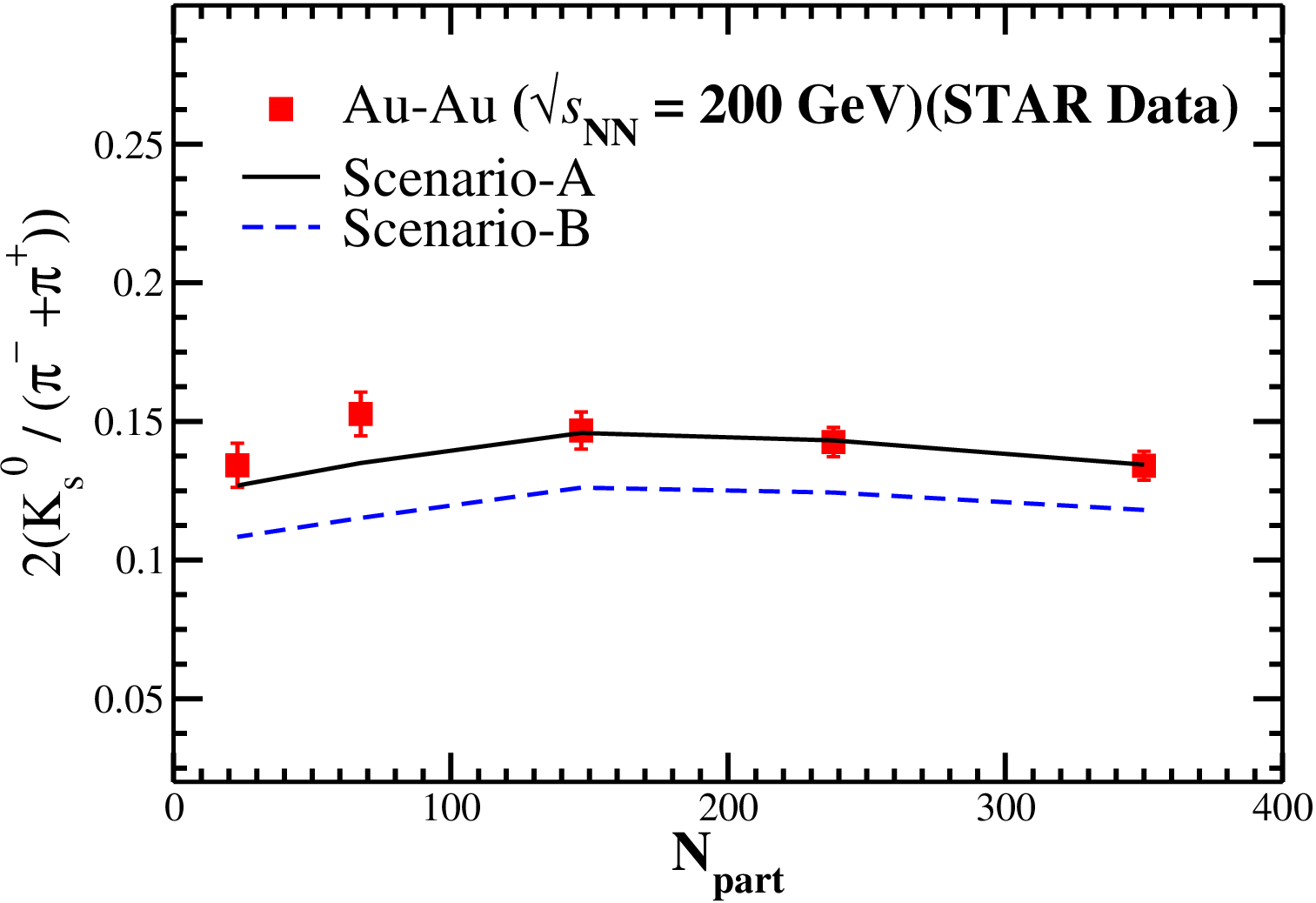}
 }
\subfloat{
   \includegraphics[width=0.44\textwidth,height=6.5cm]{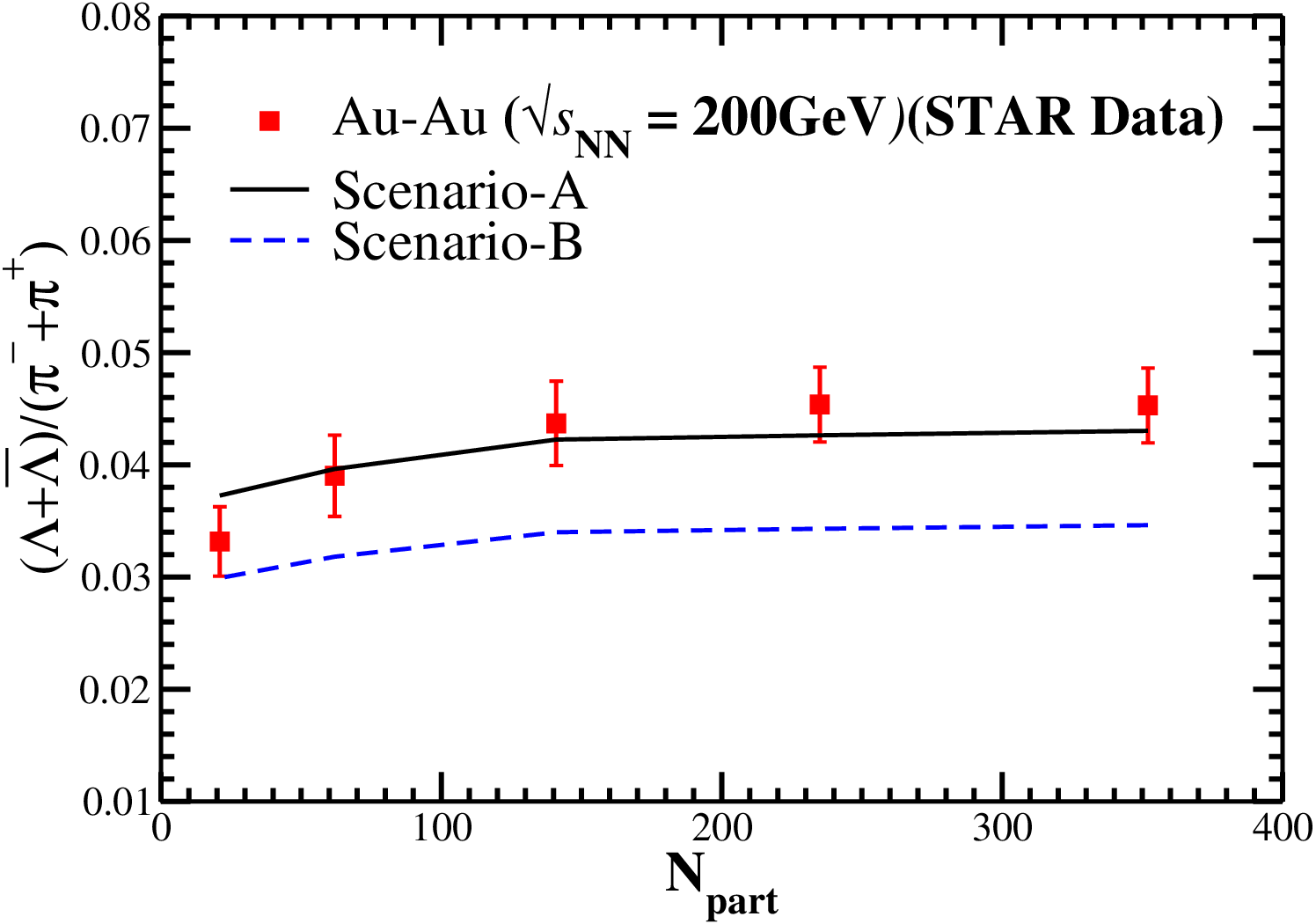}
 }
\caption{Left panel: Yield ratio for $K_s^0$ from 200 GeV Au+Au collisions. Solid points with error bars represent data points measured by STAR Collaboration as reported in\cite{agakishiev_star_prl12} and \cite{abelev_star_prc09}. The solid line is the theoretical results. Right panel: Yield ratio for $\Lambda$ from 200 GeV Au+Au collisions. Solid points with error bars represent data points measured by STAR Collaboration as reported in\cite{adam_star_prl07} and \cite{abelev_star_prc09}. The solid line is the theoretical results. }
\label{fig-hyperons-1}
\end{figure*}
Before presenting our theoretical results, we briefly discuss about the available data from Au-Au collisions at $\sqrt{s_{NN}}=200$ GeV on {\it yield ratio} based on  Ref.~\cite{adam_star_prl07} and Ref.~\cite{agakishiev_star_prl12}. For $K^0_s, \Lambda, \Xi$ hadrons, mid rapidity data ($|\eta| <0.5$) are available for five collision centralities; 0-5 \%, 10-20 \%, 20-40\%, 40-60\% and 60-80\%. In case of $\Omega$, the data are available in all those centralities apart from 10-20\% and 60-80\%. The $p_T$ integrated yield, $dN/dy$, have been measured for $\Lambda, \bar{\Lambda}, \Xi,\bar{\Xi}, \Omega, \bar{\Omega}$ in Ref.~\cite{adam_star_prl07} and for $K^0_s$ in Ref.~\cite{agakishiev_star_prl12}. The rapidity windows for $\Lambda (\bar{\Lambda})$ is $|y|<1$, for $\Xi^-(\bar{\Xi^+})$ is $|y|<0.75$ and for $\Omega^-+\bar{\Omega^+}$  is $|y|<0.75$. The intergated yield $dN/dy$ is dominated by low $p_T$ region of the spectra, that basically corresponds to soft production. $dN/dy$ of pions from Au-Au collisons at same energy is measured in \cite{abelev_star_prc09}. We take those experimental values and evaluate the ratios of the integrated yield of particles (plus anti-particles) to the pion yields, which are tabulated in the last columns of Tables-~\ref{table_data_k0s} \& \ref{table_data_hyperon} . It may be reminded that there are minor differences in $N_{part}$ values (at same centrality) in following  references~\cite{adam_star_prl07,abelev_star_prc09,agakishiev_star_prl12}. However the differences are very small as shown in Fig.\ref{fig_npart_centrality}. We consider $N_{part}$ from Ref.~\cite{adam_star_prl07} in our calculation of {\it yield ratio} of hyperons. While calculating {\it yield ratio} of $K^0_s$, we consider $N_{part}$ from Ref.~\cite{agakishiev_star_prl12}. 


While evaluating the yield of $\Lambda$ we have included contribution from $\Sigma$. Because we know it is difficult to isolate $\Lambda^0$ from $\Sigma^0$, as $\Sigma^0$ decays to $\Lambda^0$ and $\gamma$ through a weak process having 99\% branching ratio. This makes the reconstruction of $\Sigma^0$ a difficult task. Hence, data of $\Lambda$ contains $\Sigma$. The data takes into account the feed down corrections from $\Xi$, but not from $\Omega$, $\Sigma^*(1385)$ family ($\Sigma^{+*}, \Sigma^{0*}, \Sigma^{-*}$) and $\Sigma^*(1660)$. In order to take care of the feed down contribution in our calculations, we multiply a constant factor 0.8 to the $\Lambda$ yield. 

The $p_T$ window ($0.5 \text{ GeV}< p_T  < 4.8 \text{ GeV}$) in which the data of integrated yields are presented covers the domain of soft productions which is within the scope of present theoretical calculation. We present the results in Fig.\ref{fig-hyperons-1} that displays the {\it yield ratio} of $K^0_s$ and $\Lambda$. We have considered 2$K^0_s=K^++K^-$. In both left and right panels, the solid points represent the data from Au-Au collisions by STAR Collaboration\cite{adam_star_prl07,agakishiev_star_prl12,abelev_star_prc09} and the solid curves are our numerical results. Initial number density is taken as $n_i$=0.85$n_{eq}$, where $n_{eq}$ denotes the equilibrium density at $T_i$. This is described as scenario-A. We tabulate the values of $T_{ch}$ in Table~\ref{table_initialcondition} that best explains the data in each centrality. 
We redo our calculations with different initial number density, $n_i=0.7 n_{eq}$, keeping $T_{ch}$ same (Scenario B in Fig.\ref{fig-hyperons-1}). The fitting is not well as it is displayed; the lowest data point is explained well ($\Lambda$) while the other points are underpredicted. We checked with other combination of parameters but the fitting is not good. 

\begin{figure*}
\centering
\subfloat{
   \includegraphics[width=0.44\textwidth,height=6.5cm]{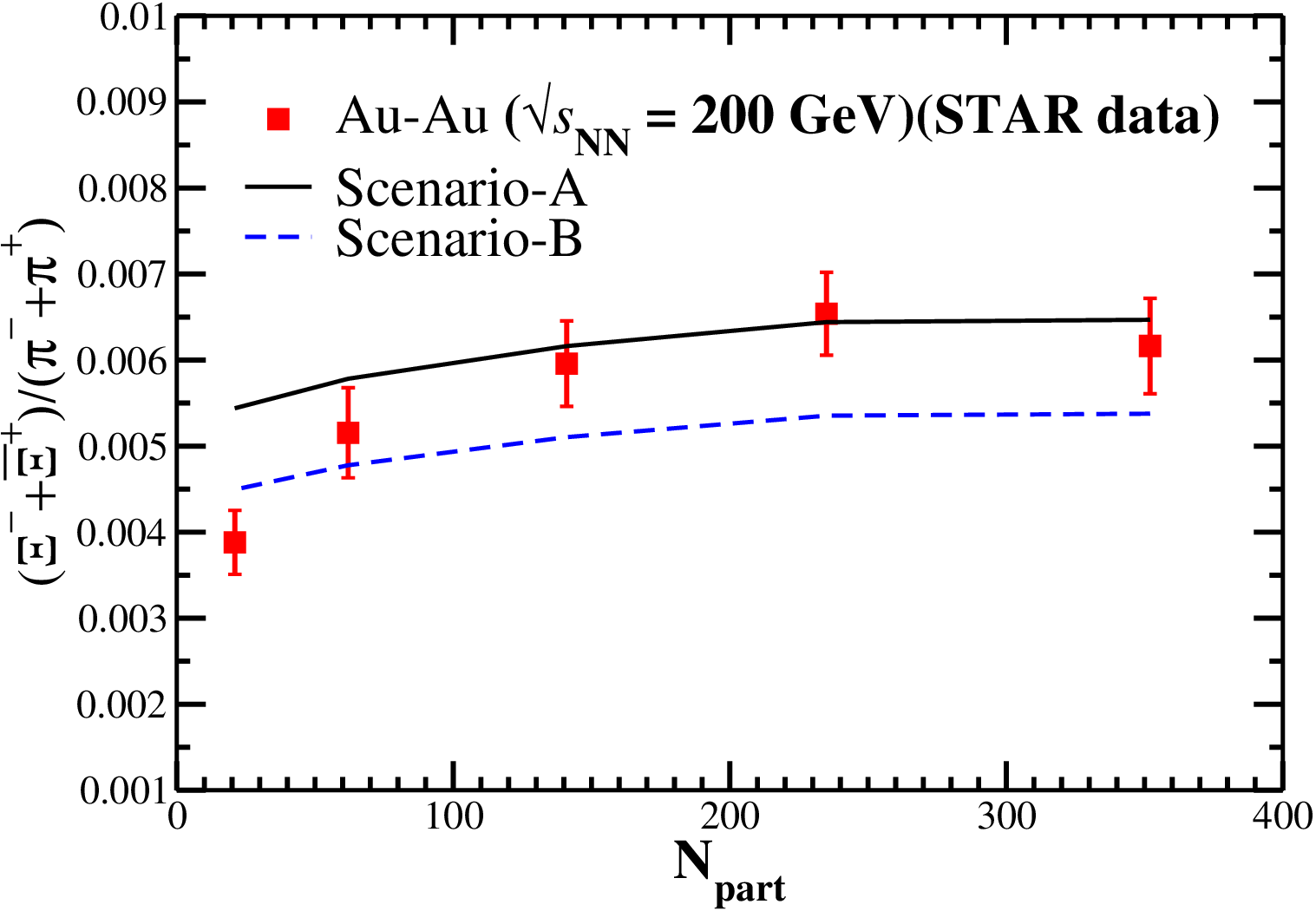}
 }
\subfloat{
   \includegraphics[width=0.44\textwidth,height=6.5cm]{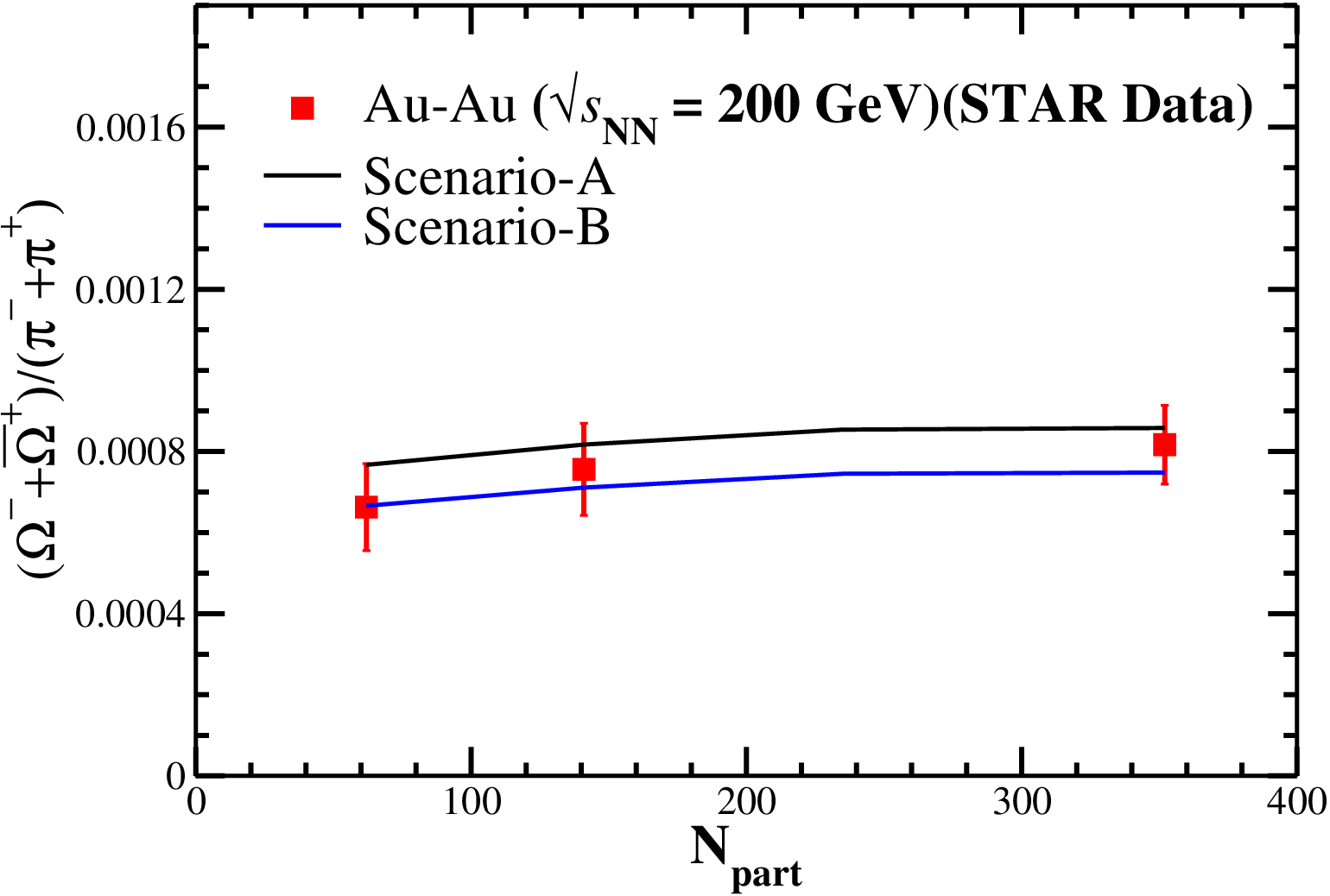}
 }
\caption{Left panel: Yield ratio for $\Xi$ from 200 GeV Au+Au collisions. Solid points with error bars represent data points measured by STAR Collaboration. The solid line is the theoretical results. Right panel: Yield ratio for $\Omega$ from 200 GeV Au+Au collisions. Solid points with error bars represent data points measured by STAR Collaboration. The solid line is the theoretical results. Data are taken from\cite{adam_star_prl07} and \cite{abelev_star_prc09}}.
\label{fig-hyperons-2}
\end{figure*}
The {\it yield ratio} for multi-strange baryons, $\Xi$ and $\Omega$, are displayed in Fig.\ref{fig-hyperons-2} for both the scenarios A \& B. The solid points of both panels represent the data from \cite{adam_star_prl07,abelev_star_prc09}, while the solid curves are our results with the same initial conditions considered to explain $K^0_s$ and $\Lambda$ in Fig.\ref{fig-hyperons-1}. Scenario-A with $n_i=0.85 n_{eq}$ explains the data points nicely except for the most peripheral one (having centrality, 60-80 \%). In case of $\Omega$ the datum point is not available for 60-80\% centrality, but it is also expected that scenario A would over-predict. However, a lower initial number density ($n_i<0.7n_{eq}$) is observed to explain $\Xi, \Omega$ yield ratio at most peripheral collsion, which is also expected from scenario B. Here also, we observe a steep rise in the {\it yield ratio}, when we move from (60-80)\% centrality to (20-40)\% centrality. For both scenarios, A and B, the $T_{ch}$ are kept same as mentioned in Table-\ref{table_initialcondition}. One can explore other scenarios with different initial conditions. Similar analysis has been carried out by us while explaining the data at LHC~\cite{ghoshprd20} at a different collision energy. From the analysis at both colliding energies we find a similar trend {\it i.e,}  lower initial densities explain better the most peripheral collisions. This is expected as a system with smaller energy densities might be produced at most peripheral collisions; which produce less particle densities.

\begin{table*}
 \caption{Parameters with freeze-out temperature($T_{ch}$, that explains the data. Reference Figs.\ref{fig-hyperons-1},\ref{fig-hyperons-2},\ref{fig_relativeyield}. }
 \centering
 \begin{tabular}{|c|c|c|c|c|c| }
   \hline
   $N_{part}$&${C_s}^{2}$ & $T_{ch}(K_s^0)$ & $ T_{ch}(\Lambda)$ & $T_{ch}(\Xi)$ & $T_{ch}(\Omega)$  \\
   & & (in GeV) & (in GeV) & (in GeV) & (in GeV)  \\
   \hline
   352  & 1/5 & 148.3 & 152 & 154.5 &155\\
   \hline
   235  & 1/5 & 149.3 & 151 & 154.2&*\\
   \hline
   141 & 1/5 & 149.2 & 150 &  150  & 150.19    \\ 
  \hline
   62 & 1/5  & 144.5  & 144.5 & 144.5 & 145    \\
  \hline
   21 & 1/5 & 139 & 139 & 139  & * \\
  \hline
 \end{tabular}
 \label{table_initialcondition} 
 \end{table*}
\begin{table*}
 \centering 
 \caption{Relative yield of anti-hyperons per participant at various centralities (with respect to most peripheral collision) from Au+Au collisions at $\sqrt{s_{NN}}$=200 GeV. Relative $(dN/dy)/N_{part}$=$\frac{[(dN/dy)/N_{part}]_{x}}{[(dN/dy)/N_{part}]_{60-80\%}}$, where $x$ represents centrality.  }
 \begin{tabular}{ |c|c|c|c|} 
  \hline
  Centrality \% & $N_{part}$ & Relative $(dN/dy)/N_{part}$ $(\bar\Lambda)$ & Relative $(dN/dy)/N_{part}$ $(\bar\Xi^+)$ \\
    \hline
   0-5 & 352 $\pm$3  & 1.578 $\pm$ 0.476 & 1.7897 $\pm$0.565 \\
  \hline
   10-20 & 235$\pm$9 & 1.433$\pm$0.434  & 1.67$\pm$0.513  \\ 
   \hline
   20-40 & 141$\pm$8 & 1.334 $\pm$ 0.398 & 1.513$\pm$0.463  \\
  \hline
   40-60 & 62$\pm$9  & 1.157$\pm$0.387  & 1.277 $\pm$ 0.437  \\
  \hline
   60-80 & 21$\pm$6 & 1.0$\pm$ 0.414 & 1.0$\pm$0.417  \\
  \hline
 \end{tabular}
 \label{table_data_relatve_yield} 
\end{table*}

\begin{figure}
\includegraphics[width=0.44\textwidth,height=6.5cm]{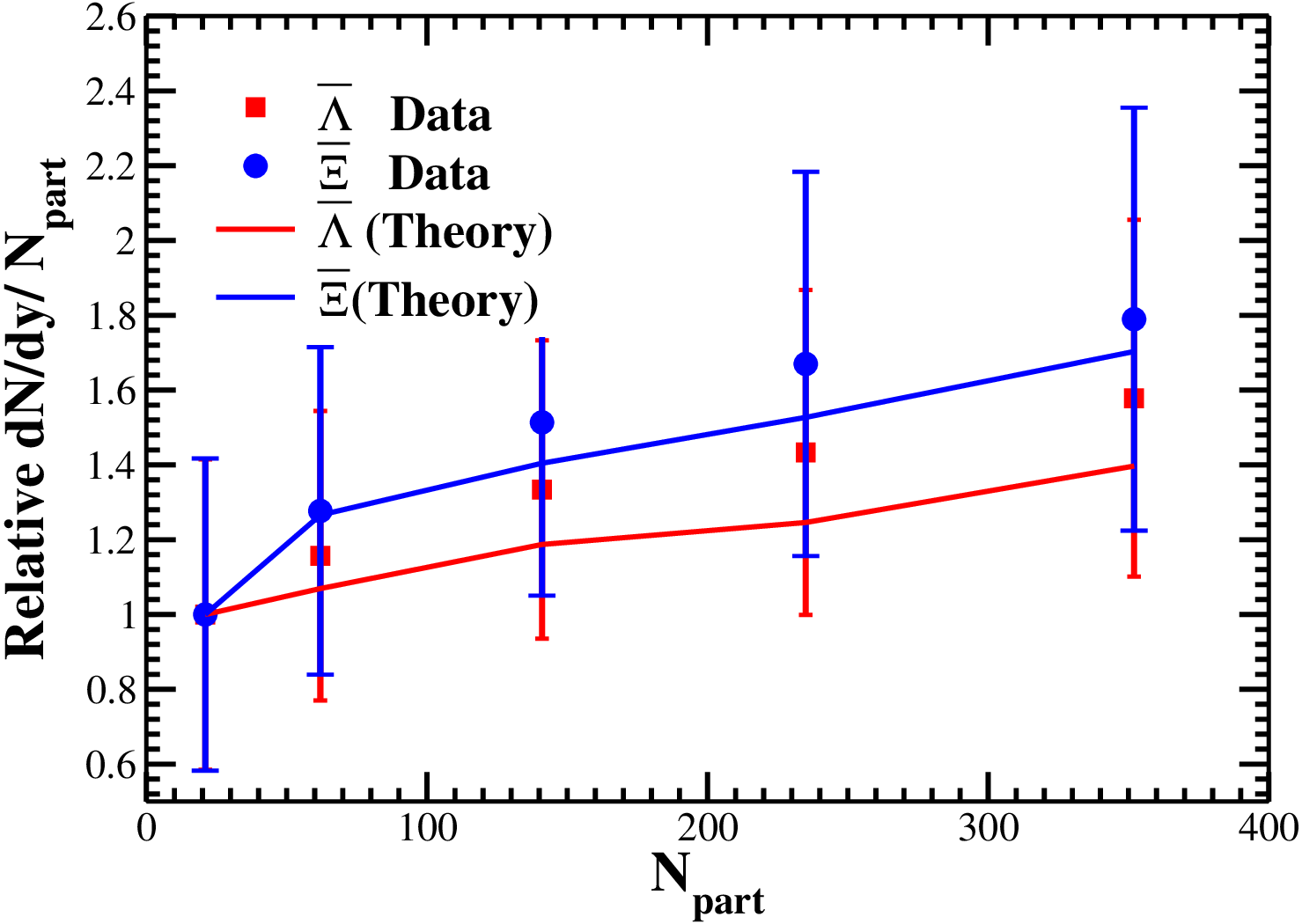}
\caption{ Relative $dN/dy/N_{part}$ is basically the ratio of yield/$N_{part}$ at certain centrality to the yield/$N_{part}$ at most peripheral(60-80\%). The red color is for the $\bar{\Lambda}$ and blue for $\bar\Xi$ hyperons. The error bar includes both statistical and systematic.}
\label{fig_relativeyield}
\end{figure}

Let's discuss the other observable, relative yield of strange hadrons. It may be noted that the enhancement of multi strange productions is highlighted in \cite{adam_star_prl07} by measuring the relative yield. It is basically the normalized yield in central collision to the normalized yield at most peripheral collision. The yield is normalized with $N_{part}$. The relative yield is written as  $dN/dy/N_{part}$ (= $((dN/dy)/N_{part})_{centrality}/((dN/dy)/N_{part})_{most peripheral}$). When the comparsion of relative yield is made for multistrange($\bar{\Xi^+}$) anti hyperon with single strange ($\bar\Lambda$) anti hyperon, then it is observed the value of the relative yield is more for multi strange compared to the single strange hadron. The values of the {\it relative yields} of anti-hyperons $\bar{\Lambda}$ and $\bar{\Xi^+}$ are tabulated in Table~\ref{table_data_relatve_yield}. Fig.\ref{fig_relativeyield} shows the relative yield per participant with $N_{part}$. The relative yield per participant is presented with respect to most peripheral collisions i.e., centrality (60-80)\%. In a particular centrality class $x$, relative yield is $(dN/dy)/N_{part}$=$[(dN/dy)/N_{part}]_{x}/[(dN/dy)/N_{part}]_{60-80\%}$ \cite{adam_star_prl07}. 
The blue solid points are data points for $\bar\Xi$ and red points are for $\bar\Lambda$. Solid lines are numerical results. It is worth to mention here, that the data we consider from \cite{adam_star_prl07} (Fig.2 of the paper) displays the statistical error bars only. Here we consider both statistical and systematic error. Both for $\bar\Lambda$ and $\bar\Xi$, there is an increase in yield per participant in central collisions compared to peripheral collisions. The increase is more in case of $\bar\Xi$ compared to $\bar\Lambda$. 

\begin{figure}[t]
\includegraphics[width=0.44\textwidth,height=6.5cm]{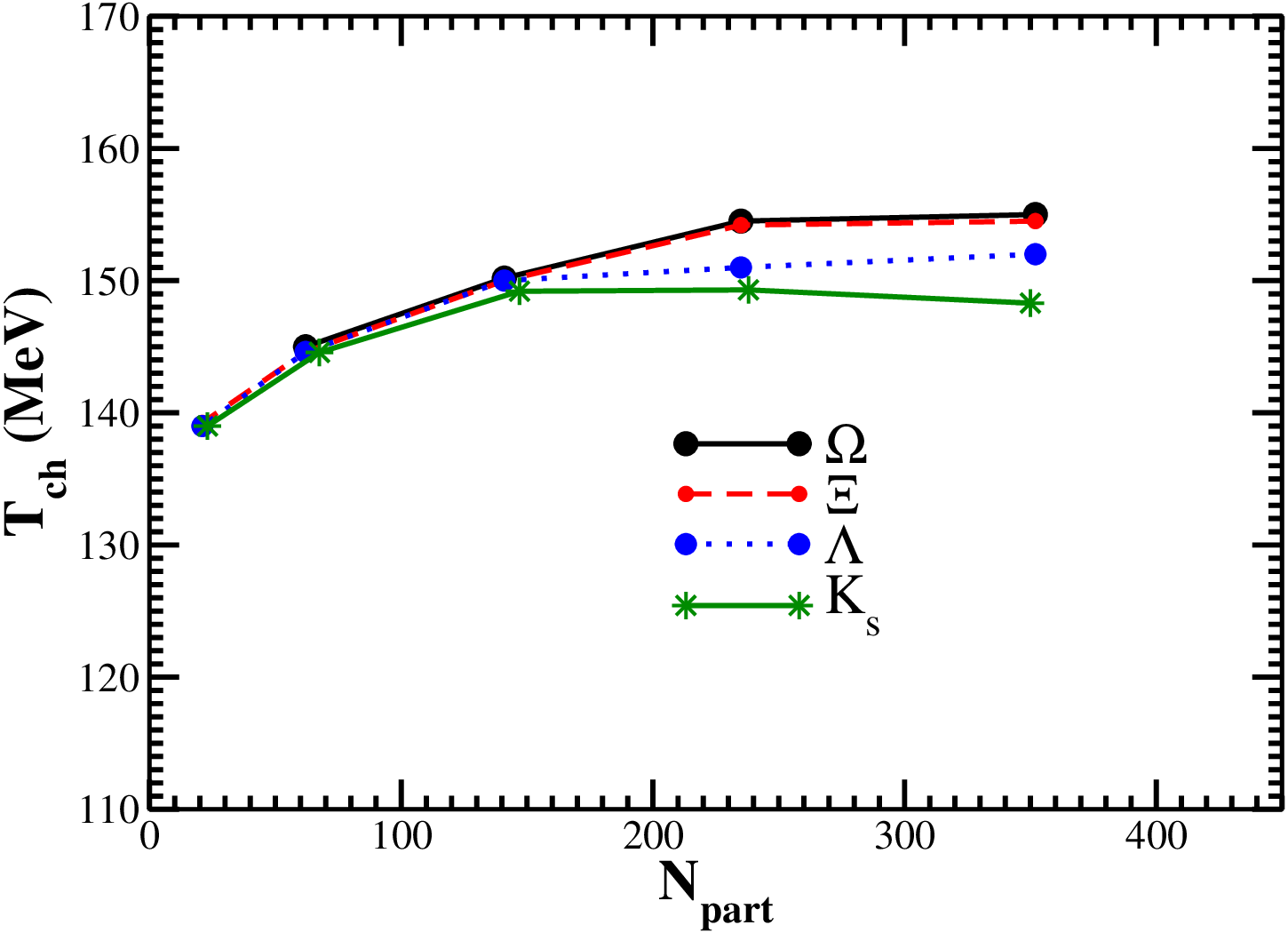}
\caption{The corresponding $T_{ch}$ for hyperons and strange mesons, that explain the above data.}
\label{fig_freezeouttemp}
\end{figure}

The freeze-out temperatures($T_{ch}$) considered in our calculation to explain simultaneously the {\it yield ratio} in Figs.\ref{fig-hyperons-1},\ref{fig-hyperons-2} and {\it relative yields} \ref{fig_relativeyield} are plotted in Fig.\ref{fig_freezeouttemp}. In case of central collsions, multi strange hadrons $\Xi, \Omega$ freeze out at a temperature close to very close to $T_c$. $\Lambda$ freezes out little later. $K^0_s$ freezes out later. In peripheral collisions all freeze out simultaneously showing a common freeze-out temperature. It is also observed that $T_{ch}$ decreases with $N_{part}$. 

\section{\label{sec:summary} Summary and Conclusions}
Multi strange productions from Au+Au collisions at $\sqrt{s_{NN}}$=200 GeV have been discussed microscopically using rate equations considering the cross sections of various hadronic processes. Rate equations are solved simultaneously both for single strange $K, \Lambda, \Sigma$ and multi strange $\Xi$ and $\Omega$ hadrons. The yield of strange hadrons with their anti-particles are normalised with thermal pions to get the {\it yield ratios}, such as, $(K^++K^-)/(\pi^++\pi^-)$, $(\Lambda+\bar{\Lambda})/(\pi^++\pi^-)$, $(\Xi^-+\bar{\Xi^+})/(\pi^++\pi^-)$ and $(\Omega^-+\bar{\Omega^+})/(\pi^++\pi^-)$ at various centralities. The results are then compared with the experimental data from Refs.~\cite{adam_star_prl07,agakishiev_star_prl12}. The {\it yield per participant} for anti-hyperons $\bar\Xi, \bar\Lambda$ are also evaluated for various $N_{part}$ and compared with experimental observations. The freeze out temperatures are also extracted for various centrality of collisions. From the results of {\it yield ratios} of $K^0_s$, $\Lambda$, $\Xi$ and $\Omega$, it has been observed that the data are explained satisfactorily with $n_i=0.85n_{eq}$ for central collisions where $N_{part}$ is large. For most peripheral collisions, the hyperon productions are not well explained with $n_i=0.85 n_{eq}$, rather these are explained with less initial densities, $n_i =0.7 n_{eq}$. A lower value of initial number density in most peripheral collision may indicate that less initial energy densities are created in most peripheral collisions due to less number of participating nucleons, as expected.

The rise of {\it yield ratios} with $N_{part}$ is observed for all strange hadrons which indicates the increase in strangeness production with centrality. Large steepness is seen in case of multi-strange hadrons compared to $K$ and $\Lambda$. This tells about the enhancement of multistrange production rate. This probably says that the probability of formation of multistrange hadrons from the bulk increases with $N_{part}$ due to increase in number of interactions. It may also be argued in another way, that there is an increase in the rate of more massive hadron productions with centrality (as steepness is more for heavier Cascade compared to Kaon and Lambda), indicating some sort of mass ordering. 

The above argument is again more pronounced from the observation of {\it yield per participant} with $N_{part}$. There is a systematic increase in the {\it yield per participant} with $N_{part}$ as one moves from lighter proton to heavier Cascade \cite{adam_star_prl07}. Instead of attributing this phenomenon to an increase in the number of initial constituents, this may be attributed to mass ordering. However, the $R_{CP}$ measurements of $\Lambda,\Xi,\Omega$ with $p_T$ in \cite{adam_star_prl07} discards the reason related to mass, but attribute the same to the constituent quarks, the density of which increases with centrality. In this theoretical calculation, we observed an increase in initial number densities of strange hadrons from most peripheral ($n_i=0.7n_{eq}$) to central ($0.8n_{eq}$) collisions, which is required to explain the data. Initial number densities are parameteres in this study, but in principle should be constrained from strange quark densities assuming an initial QGP phase. This is kept for a future work. 

The {\it relative yield per participant} basically tells about the scaling property of baryons(anti baryons) with centrality. Here it is shown by considering anti-baryons $\bar\Lambda$ and $\bar\Xi$. Strange anti-baryons are chosen as scaling observables because the valence quark content of these particles can only be created in the collisions. Hence, the information from this observable can shed light on the produced system. Even if one considers the {\it relative yield per participant} with strange baryons, the inference would be similar at this colliding energy. This is because baryon to anti-baryon ratio is close to one at such high energy collisions. The analysis is done only for $\bar\Lambda$ and $\bar\Xi$ as data of $\bar{\Omega}$ are not available. 

It may be pointed out that similar behaviour of {\it yield ratio}, $(H_s+\bar{H_s})/(\pi^++\pi^-)$ is observed from Pb-Pb collisions at $\sqrt{s_{NN}}$=2.76 TeV, LHC energy \cite{ghoshprd20}. The multistrange data at most peripheral collision are over predicted when $n_i$ for both central collision and most peripheral collision are considered same. The system that is produced in most peripheral collisions may be far away from thermodynamic equilibrium and should not be treated at par with most central collisions at RHIC, $\sqrt{s_{NN}}$=200 GeV, and at LHC, $\sqrt{s_{NN}}$=2.76 TeV.   While moving from most peripheral to central collisions, when a smooth behaviour of the {\it yield ratio} is obtained instead of steep rise, one probably expect a change in the medium property. Now question arises, whether the change in medium property can be attributed to (i) the presence or abscence of initial partonic system or (ii) presence of initial partonic system but with less or more strange parton density? It is a matter of future investigation.

The freeze-out temperatures obtained within the ambit of this present calculation signify a sequential freeze out for multi strange \& single strange hadrons in case of central collsions and a common freeze out for all in most peripheral collisions. Multi strange $\Xi, \Omega$ freeze out at a temperature very close to $T_c$. It is also observed that $T_{ch}$ decreases with $N_{part}$. 
\onecolumngrid
\appendix*
\section{Cross-section}
 The cross sections for various hadronic processes producing hyperons and strange mesons are listed below along with corresponding references.
 The cross-sections of all inverse reactions are obtained using principle of detailed balance as follows; 
$\sigma_{f\rightarrow i}=\frac{{P_i}^2}{{P_f}^2}\frac{g_i}{g_f}\sigma_{i\rightarrow f}$
where $P_i, P_f$ are the center of mass momenta and $g_i, g_f$ are the total degeneracies of the initial and final channels. 
 \begin{table}[H] 
 \caption{Cross sections for $K,\, \Lambda,\, \Sigma$ production}
 \label{tab:kproduction_ch}
 \centering
 \begin{tabular}{ |>{\centering\arraybackslash}m{2.7cm}|>{\centering\arraybackslash}m{8cm}|>{\centering\arraybackslash}m{6cm}| }
  \hline
  Reaction & Cross-section & Remarks\\
  \hline
  \multirow{10}{*}{\shortstack{$\pi\pi\rightarrow K\bar{K}$~\cite{Brown1}\\$\rho\rho\rightarrow K\bar{K}$\\ $\pi\rho\rightarrow K\bar{K^*}$ \\ $\pi\rho\rightarrow {K^*}\bar{K}$}} & \multirow{10}{6cm}{${\bar{\sigma}}_{ab\rightarrow cd}(s)=\frac{1}{32 \pi} \frac{P'_{cd}}{sP_{ab}} \int^1_{-1}dx M(s,x)$} & \multirow{10}{6cm}{$P_{ab},\, P'_{cd}$ - three momenta of incoming mesons and outgoing Kaons in the CM frame, $x= \text{cos}(P_{ab},P'_{cd})$. $M(s,x)$ is the invariant amplitude calculated using following interaction Lagrangian densities~\cite{Brown1}
  $${\mathcal L}_{K^*K\pi} = g_{K^*K\pi}K^{*\mu}\tau[K(\partial_{\mu}\pi)-(\partial_{\mu}K)\pi]$$
  $${\mathcal L}_{\rho KK} = g_{\rho KK}[K\tau(\partial_{\mu}K)-(\partial^{\mu}K)\tau K]\rho^{\mu}$$}\\
   & & \\
   & & \\
   & & \\
  & &\\
  & &\\
  & &\\
  & &\\
  & &\\
  & &\\  
  \hline
  \multirow{9}{*}{\shortstack{$\pi N \rightarrow \Lambda K$~\cite{Brown1,cugnonnpa84} \\ $\pi N \rightarrow \Sigma K$ \\ $\rho N \rightarrow \Lambda K$ \\$\rho N \rightarrow \Sigma K$}} & \multirow{9}{*}{\shortstack{$\bar{\sigma}_{MB\rightarrow YK} = $ \\ $\sum_i\frac{(2J_i +1)}{(2S_1+1)(2S_2+1)} \frac{4\pi}{k_i^2}\frac{\Gamma_i^2/4} {(\sqrt{s}-m_i)^2+\Gamma_i^2/4} B_i^{\text{in}} B_i^{\text{out}}$\\Here $M$ and $B$ stands for Meson and Baryon resp. \\ $Y$ stands for  $\Lambda \text{ or }  \Sigma$}} & \multirow{9}{6cm}{The sum is over resonances with mass($m_i$), spin ($J_i$) and decay width($\Gamma_i$). $N^*_1(1650), N^*_2(1710), N^*_3(1720)$ \cite{amslar08} and $N^*_5(1875), N^*_6(1900)$\cite{patrignani16} are considered as intermediate resonant states. $(2S_1+1)$ and $(2S_2+1)$ are the polarisation states of the meson ($M$) and baryon($B$) in the incoming channels. $B_i$ represents the branching ratio.} \\
  & & \\
  & & \\
  & & \\
  & & \\
  & & \\
  & & \\
  & & \\
  & & \\
  \hline
  \multirow{7}{*}{${\bar K}N\rightarrow \Sigma \pi$~\cite{linpa97,ko83plb}} & \multirow{7}{*}{\shortstack{\footnotesize $\sigma_{{\bar K}N\rightarrow \Sigma \pi}=\sigma_{{K^-}p\rightarrow \Sigma^0 \pi^0}+\sigma_{{K^-}n\rightarrow \Sigma^0 \pi^-}$, \\ \footnotesize
  where  $\sigma_{{K^-}p\rightarrow \Sigma^0 \pi^0}\approx\sigma_{{K^-}n\rightarrow \Sigma^0 \pi^-}$ and \\ \tiny 
  $\sigma_{{K^-}p\rightarrow \Sigma^0 \pi^0} = \begin{cases}
                                              0.624~p^{-1.83}\text{ mb} & \text{if} ~~p \leq 0.345 \text{ GeV} \\
                                              \\
                                              \dfrac{0.0138}{(p-0.385)^2+0.0017} \text{ mb} & \text{if}~~ 0.345< p \leq 0.425 \text{ GeV}\\
                                              \\
                                              0.7~p^{-2.09}\text{ mb} & \text{if}~~p> 0.425 \text{ GeV}
                                             \end{cases}$ }} & \multirow{7}{6cm}{$p$ is the anti-Kaon momentum in the Lab frame}\\
  & & \\
  & & \\
  & & \\  
  & & \\
  & & \\   
  & & \\   
  \hline
  \multirow{6}{*}{$\bar{K}N\rightarrow \Lambda \pi$~\cite{linpa97,ko83plb}} & \multirow{6}{*}{$\sigma_{{K^-}p\rightarrow \Lambda \pi^0} = \begin{cases}
                                             1.205~p^{-1.428}\text{ mb} & \text{if} ~~p \geq 0.6 \text{ GeV}\\
                                             3.5~p^{0.659}\text{ mb} & \text{if}~~ 0.6 < p \leq 1.0 \text{ GeV}\\
                                             3.5~p^{-3.97}\text{ mb} & \text{if}~~p> 1.0 \text{ GeV}
                                            \end{cases}$ } & \multirow{6}{6cm}{$p$ is the anti-Kaon momentum in the Lab frame}\\
  & & \\
  & & \\
  & & \\  
  & & \\
  & & \\     
  \hline
  \multirow{6}{*}{\shortstack{$pp\rightarrow K\bar{K}$~\cite{kaidalov94,titov08,boreskov83} \\ $pp\rightarrow \Lambda\bar{\Lambda}$\\ $pp\rightarrow \Sigma\bar{\Sigma}$}} & \multirow{6}{*}{\shortstack{ $\sigma_{P \bar{P}\rightarrow \bar{Y}Y(\bar{K}K)}=
 \frac{C_AC_{Y_{i}(K)}g_{0}^4}{16\pi}\times\frac{s}{s-4m_{P}^{2}}\times$ \\ $\Gamma\left(1-\alpha(0)\right)^2\times
 \left(\frac{s}{s_{0}^{\bar{P}P\rightarrow \bar{Y} Y(\bar{K}K)}}\right)^{2(\alpha_{0}-1)}\times \frac{e^{\varLambda_1 t_{\text{min}}}}{\varLambda_1}$ }} & \multirow{6}{*}{Refer Table~\ref{table_paramet_pp} for values of parameters.} \\
  & & \\
  & & \\
  & & \\  
  & & \\
  & & \\
  \hline
 \end{tabular}
 \end{table}

 \begin{table}[H] 
 \caption{Cross sections for $\Xi$ production}
 \label{tab:xiproduction_ch}
 \centering
 \begin{tabular}{ |>{\centering\arraybackslash}m{2.7cm}|>{\centering\arraybackslash}m{8cm}|>{\centering\arraybackslash}m{6cm}| }
  \hline
  Reaction & Cross-section & Remarks\\
  \hline
  $\bar{K}\Lambda\rightarrow \pi \Xi$~\cite{chen04,linpa02} & $\sigma_{\bar{K}\Lambda\rightarrow \pi \Xi} = \dfrac{1}{4}\dfrac{p_{\pi}}{p_{\bar{K}}}\mid M_{\bar{K}\Lambda\rightarrow \pi \Xi}\mid^{2}$ &  $p_\pi$ and $p_K$ - momentum in CM frame\\
   & with $\mid M_{\bar{K}\Lambda\rightarrow \pi \Xi}\mid^{2}=34.7~\dfrac{s_0}{s}$ & $s_0 = \sum_i m_i$ - threshold energy\\
  \hline
  $\bar{K}\Sigma\rightarrow \pi \Xi $~\cite{chen04,linpa02} & $\sigma_{\bar{K}\Sigma\rightarrow \pi \Xi} = \dfrac{1}{12}\dfrac{p_{\pi}}{p_{\bar{K}}}\mid M_{\bar{K}\Sigma\rightarrow \pi \Xi}\mid^{2}$ & $p_\pi$ and $p_K$ - momentum in CM frame\\
  & with $\mid M_{\bar{K}\Sigma\rightarrow \pi \Xi}\mid^{2}=318\left( 1-\dfrac{s_0}{s}\right)^{0.6}\times \left(\dfrac{s_0}{s}\right)^{1.7}$ & $s_0 = \sum_i m_i$ - threshold energy\\
  \hline
  $\Lambda\Lambda\rightarrow N\Xi$~\cite{leeprc12,holzenkamp89,adelseck90} & $\sigma_{\Lambda\Lambda\rightarrow N\Xi}=37.15\frac{p_N}{p_\Lambda}\left(\sqrt{s}-\sqrt{s_0}\right)^{-0.16} ~~\text{mb}$ & Parametrisation valid for $0<(\sqrt{s}-\sqrt{s_0})<0.6$ GeV\\
  \hline
  $\Lambda\Sigma\rightarrow N\Xi$~\cite{leeprc12,holzenkamp89,adelseck90} & $\sigma_{\Lambda\Sigma\rightarrow N\Xi}=25.12\left(\sqrt{s}-\sqrt{s_0}\right)^{-0.42} ~~\text{mb}$ & Parametrisation valid for $0<(\sqrt{s}-\sqrt{s_0})<0.6$ GeV\\
  \hline
  $\Sigma\Sigma\rightarrow N\Xi$~\cite{leeprc12,holzenkamp89,adelseck90}  & $\sigma_{\Sigma\Sigma\rightarrow N\Xi}=8.51\left(\sqrt{s}-\sqrt{s_0}\right)^{-0.395} ~~\text{mb}$ & Parametrisation valid for $0<(\sqrt{s}-\sqrt{s_0})<0.6$ GeV\\
  \hline
  $\bar{K}N\rightarrow K\Xi$~\cite{leeprc12,holzenkamp89,adelseck90,sharov11} & $\sigma_{\bar{K}N \rightarrow K\Xi}=0.5\left[\sigma_{K^-p \rightarrow K^+\Xi^-}+\sigma_{K^-p \rightarrow K^0\Xi^0}+\sigma_{K^-n \rightarrow K^0\Xi^-}\right]$ & \\
  & with $\sigma_{K^-p \rightarrow K^+\Xi^-} =235.6\left(1-\frac{\sqrt{s_0}}{\sqrt{s}}\right)^{2.4}\left(\frac{\sqrt{s_0}}{\sqrt{s}}\right)^{16.6} \text{ mb}$ & Parametrisation valid for $0<(\sqrt{s}-\sqrt{s_0})<1$ GeV\\
  & $\sigma_{K^-p \rightarrow K^0\Xi^0}=7739.9\left(1-\frac{\sqrt{s_0}}{\sqrt{s}}\right)^{3.8}\left(\frac{\sqrt{s_0}}{\sqrt{s}}\right)^{26.5} \text{ mb}$ & \\
  & $\sigma_{K^-n \rightarrow K^0\Xi^-}=235.6\left(1-\frac{\sqrt{s_0}}{\sqrt{s}}\right)^{2.4}\left(\frac{\sqrt{s_0}}{\sqrt{s}}\right)^{16.6} \text{ mb}$ & \\
  \hline
  $\bar{p}p \rightarrow \overline{\Xi}^0\Xi^0$~\cite{kaidalov94} & $\sigma_{\bar{p}p \rightarrow \overline{\Xi}^0\Xi^0}=\frac{16}{81\pi}\frac{[\sigma_{\bar{p}p\rightarrow \bar{\Lambda}\Lambda}]^2}{2\Lambda_1}\text{exp}\left[\Lambda_1 t_{DC}\right]$ &\\
  & Here $\Lambda_1 = 9\text{ GeV}^{-2}$ & Refer Table~\ref{tab:kproduction_ch} for $\sigma_{\bar{p}p\rightarrow \bar{\Lambda}\Lambda}$\\
  $\bar{p}p \rightarrow \overline{\Xi}^+\Xi^-$~\cite{kaidalov94} &  $\sigma_{\bar{p}p \rightarrow \overline{\Xi}^+\Xi^-}=16\sigma_{\bar{p}p\rightarrow \overline{\Xi}^0\Xi^0}$ & \\
  \hline
 \end{tabular}
 \end{table}

 \begin{table}[H] 
 \caption{Cross sections for $\Omega$ production}
 \label{tab:omgproduction_ch}
 \centering
 \begin{tabular}{ |>{\centering\arraybackslash}m{2.7cm}|>{\centering\arraybackslash}m{8cm}|>{\centering\arraybackslash}m{6cm}| }
  \hline
  Reaction & Cross-section & Remarks\\
  \hline
  $K^{-} \Lambda \rightarrow \Omega^{-} K^{0}$~\cite{ghosharxiv19,gaitanos16} & $\sigma_{K^{-} \Lambda \rightarrow \Omega^{-} K^{0}}= a_0+a_1 \, p_{\text{lab}}+a_2\,  p_{\text{lab}}^2+a_3 \, \text{exp}(-a_4 p_{\text{lab}})$ & $a_0 = 0.155591$, $a_1 = -0.0473326$,  $a_2 = 0.00362302$ \\
  & $1.011\leq p_{\text{lab}} (GeV)\leq 6.55$ &  $a_3 = -0.29776$, $a_4 = 0.917116$\\ 
  \hline
  $K^{-} \Sigma^0 \rightarrow \Omega^{-} K^{0}$~\cite{ghosharxiv19,gaitanos16} & $\sigma_{K^{-} \Sigma^0 \rightarrow \Omega^{-} K^{0}}= b_0+b_1 p_{\text{lab}}+b_2 p_{\text{lab}}^2+b_3 \, \text{exp}(-b_4 p_{\text{lab}})$ & $b_0 = 0.137027$, $b_1 = -0.0422865$,  $b_2 =0.00327658$ \\
  & $1.19\leq p_{\text{lab}} (GeV)\leq 5.991$ &  $b_3 = -0.281588$, $b_4 = 0.942457$\\
  \hline
  $\pi^{0} \Xi^{-} \rightarrow \Omega^{-} K^{0}$~\cite{ghosharxiv19,gaitanos16} & $\sigma_{\pi^{0} \Xi^{-} \rightarrow \Omega^{-} K^{0}} = c_0+c_1 \, p_{\text{lab}}+c_2\, p_{\text{lab}}^2+c_3/p_{\text{lab}}+$ & $c_0 = -0.414988$, $c_1 = -0.025499$,  $c_2 =0.00628967$ \\
  & \hspace{0.4in}$c_4/(p_{\text{lab}}^2)+c_5 \text{exp}(-p_{\text{lab}})$ & $c_3 = 2.1816$, $c_4 = -0.639193$, $c_5= -2.85555$\\
  & $1.033\leq p_{\text{lab}} (GeV)\leq 5.351$ & \\
  \hline
  $\bar{p} p\rightarrow \Omega^{-} \bar{\Omega}^+$~\cite{kaidalov94} & $\sigma_{\bar{p} p\rightarrow \Omega^{-} \bar{\Omega}^+}=
 \frac{4^3}{\pi^2}\times\frac{[\sigma_{p \bar{p}\rightarrow \bar{\Lambda}\Lambda}]^3}{\Lambda_1^2}
 \times \exp[\Lambda_1 t_{DO}]$ & $t_{DO}=t_{min}^{\Lambda\Xi}-t_{min}^{p\Lambda}+t_{min}^{\Xi\Omega}-t_{min}^{p\Lambda}$\\
 & & with $t_{min}^{ij}=-\frac{s}{2}+m_{i}^{2}+m_{j}^{2}+\frac{1}{2}\sqrt{(s-4m_{i}^{2})(s-4m_{j}^{2})}$\\
 & & Refer Table~\ref{tab:kproduction_ch} for $\sigma_{\bar{p}p\rightarrow \bar{\Lambda}\Lambda}$ \\
 \hline
 \end{tabular}
 \end{table} 
\begin{table}[H]
 \caption{Parameters for $pp\rightarrow YY(MM)$ Reactions\cite{flaminio84,barnes87,hasan92,tanimori90,sugimoto88}}
 \centering
 \begin{tabular}{ |c|c|c|c|c|c| }
  \hline
   Reactions & $ C_A $ & $ C_{Y_i(M_i)} (GeV^{-2})$ & $\Lambda_1(GeV^2)$ & $s_0$ & Regge trajectory  \\
   &  &  &  & & $\alpha(t)$=  \\
  \hline
   $ pp\rightarrow K^+K^-$ & 0.08 & 4 & 4 & 1.93 & -0.86+0.5t \\   
  \hline
   $ pp\rightarrow \bar{\Lambda}\Lambda$ & 0.10 & 9/4 & 9 & 2.43 &  0.32+0.85t  \\
  \hline
   $ pp\rightarrow \bar{\Sigma^-}\Sigma^+$ & 0.10 & 1 & 9 & 2.43 & 0.32+0.85t  \\
  \hline    
 \end{tabular}
 \label{table_paramet_pp} 
\end{table}
 
\end{document}